\magnification = \magstep1
\pageno=1
\hsize=15.0truecm
\hoffset=1.0truecm
%
\vsize=23.5truecm

\output={\plainoutput}
\pretolerance=3000
\tolerance=5000
\hyphenpenalty=10000    
\newdimen\digitwidth
\setbox0=\hbox{\rm0}
\digitwidth=\wd0
\def\cl{\centerline}
\def\ni{\noindent}

\def\vs{\vskip 11pt}

\font\sc=cmr7

\def\ia{(i/\kern-2pt {\sc A})}
\def\ja{(j/\kern-2pt {\sc A})}
\vs
\cl{\bf High Resolution Irradiance Spectrum from 300 to 1000 nm}
\vs\vs
\cl{Robert L. Kurucz}
\cl{Harvard-Smithsonian Center for Astrophysics}
\cl{60 Garden Street, Cambridge, MA, USA}
\vs
\cl{June 15, 2005}
\vs
The FTS scans that made up the Kitt Peak Solar Flux Atlas by Kurucz, 
Furenlid, Brault, and Testerman 1984 have been re-reduced.  The scans 
listed in Table 1 (= Figure 0) were smoothed with a 3-point Gaussian to 
simplify continuum placement and matching overlapping scans.  An approximate 
atmospheric model was determined for each FTS scan.  Large-scale features 
produced by O$_{3}$ and O$_{2}$ dimer were computed and divided out.  The 
telluric line spectrum was computed using HITRAN and other line data 
for H$_{2}$O, O$_{2}$, and CO$_{2}$.  The line parameters were adjusted 
for an approximate match to the observed spectra.  The wavelength scale for 
the scans was redetermined.  The solar continuum level was found by
fitting a smooth curve to high points in the observed spectrum while
comparing with the product of the computed solar spectrum times the 
computed telluric spectrum.  The spectrum was normalized to the fitted 
continuum to produce a residual flux spectrum for each FTS scan.  
Those scans were divided by the computed telluric spectra to produce 
residual irradiance spectra.  Artifacts from wavelength mismatches, 
deep lines, etc, were removed by hand and replaced by linear interpolation.
Overlapping scans were fitted together to make a continuous spectrum 
from 300 to 1000 nm.  All the above steps were iterative.  The monochromatic 
error varies from 0.1 to 1.0 percent.

Given a calculated or semiempirical solar model, the continuum level 
can be computed and multiplied by the residual irradiance spectrum 
to produce the absolute irradiance spectrum at high resolution.  
Alternatively, the high resolution residual irradiance spectrum can 
be broadened and smoothed to match the resolution of any low resolution 
irradiance spectrum and normalized to the low resolution spectrum.  
That normalization can be applied to the high resolution spectrum 
to obtain a high resolution absolute irradiance spectrum.  An example 
for each method is presented below.  This is the spectrum that illuminates the 
Earth and all other bodies in the solar system.  This is a typical 
G star spectrum like those that illuminate extra-solar planets.

A revised solar flux atlas, a central intensity atlas, and a limb 
intensity atlas will be produced.  Atlases with sample computed 
spectra and line identifications will be produced.  This work, and
the extension to longer wavelengths, requires funding, as indicated below.
\vs
\vs
\vs
\hrule width 100pt
\vskip 5pt
\ni Presented at the AFRL Transmission Meeting, 15-16 June 2005, Lexington, Mass.
\vfill
\eject

\ni Figure 1 shows the old Kitt Peak Solar Flux Atlas.  
There are eight overlapping FTS scans that were normalized and pieced
together to form a continuous residual spectrum.  The observational data
for the scans is listed in Table 1.  (There are only seven scans in this 
figure because the eigth is beyond 1000 nm.)   The next figures show the
re-reduction of these scans.
\vs
\ni Figure 2 shows broad atmospheric features of O$_{3}$ and [O$_{2}$]$_{2}$ 
that were present in the scans but not considered.  Each scan was assigned to an
atmospheric model listed in Table 1.  The O$_{3}$ and [O$_{2}$]$_{2}$ transmission was 
computed using programs available on my website, kurucz.harvard.edu,   
and divided out.  (The transmissions for the seven scans were pieced together 
for the plot.)
\vs
\ni Figure 3 shows the beginning of one of the FTS scans in green.  A
continuum, smooth green line, is subjectively fitted to the scans by
comparing to predictions from calculations of the solar spectrum and 
the telluric spectrum.  When a reasonable looking fit has been obtained
through iteration, the spectrum is divided by the continuum value to
produce a residual spectrum shown in red.  The top 1 percent of the 
residual spectrum is replotted in red as well.  The blue curve is the 
the transmission curve for O$_{3}$ and [O$_{2}$]$_{2}$ that has already 
been divided out.
\vs
\ni In Figure 4 the scans were blueshifted to remove the gravitational 
red shift and pieced together in the solar laboratory frame in air.  
This is the revised spectrum of the Kitt Peak Solar Flux Atlas. 
\vs
\ni Figure 5 shows telluric lines of O$_{2}$ and H$_{2}$O that were computed 
from the atmospheric model for each scan in the solar laboratory frame
with gravitational red shift removed.  (The seven scans are pieced together.)
\vs
\ni Figure 6 shows a sample calculation of the spectrum for a relatively empty 
angstrom at 599.1 nm in the Solar Flux Atlas shown in Figure 4.  The telluric,
solar, and observed spectra are labelled at normal scale and 10 times scale.
\vs
\ni Figure 7 shows the irradiance spectrum obtained from the spectra in 
Figure 6 by dividing out the telluric spectrum.  For stronger telluric lines
and lines with incorrect wavelengths, there are artifacts that appear in the 
irradiance spectrum that were removed by hand and replaced with a linear
interpolation. 
\vs
\ni Figure 8 shows the residual irradiance spectrum after all the scans have
been processed and pieced together in the solar laboratory fram in vacuum
with gravitational red shift included.
\vs
\ni Figure 9 shows the predicted level of the continuum for theoretical solar 
model ASUN (Kurucz 1992).
\vs
\ni Figure 10 shows the absolute irradiance spectrum obtained by normalizing
the residual irradiance spectrum shown in Figure 8 to the continuum level
shown in Figure 9.
\vs
\ni Figure 11 shows the reference irradiance spectrum proposed by Thuillier et al
(2004).

\vs
\ni Figure 12 shows the Kitt Peak absolute irradiance spectrum smoothed using 
a 0.5 nm triangular bandpass that approximates the resolution of Thuillier et al 
and then compares the two spectra.  Note the
probable overestimation of the ozone below 320 nm and around 600 nm in the
Kitt Peak atlas.  (Remember that ozone has been divided out.)  I will probably 
have to re-reduce those scans.  Note the flux discrepency in the G band.  It 
appears that model ASUN does not produce enough flux, perhaps becase of 
insufficient opacity below 300 nm that results in too low a temperature gradient.
I am adding more line opacity.  I will try to produce a better model.  Of course, 
there may also be errors in Thuillier et al as well.
\vs
\ni Figure 13 shows the Kitt Peak irradiance spectrum subjectively normalized 
to the Thuillier et al irradiance spectrum.  I recommend this spectrum as the
high resolution irradiance spectrum.  The procedure for removing telluric lines
introduces noise into the irradiance spectrum where there were telluric lines.  
The flux atlas itself should be used for abundance analysis or other critical work.

\vs

\ni Funding

     NASA discontinued my funding on May 15, 2004 and I was laid off.  I 
decided to do this project using my retirement funds in order to try to 
get interest and funding from other NASA programs, or the Air Force, NOAA, 
DOE, or anyone else.  Producing the irradiance spectrum took 129 12-hour 
days at \$1000/day.  I made 115,000 11x17 plots at \$0.28/page.  I cannot
continue to subsidize NASA, or the Air Force, or NOAA with my retirement funds.
I will sell a CD and a set of plots for \$161,000 that I will return to
my retirement fund.  If I am to extend the irradiance spectrum 
into the infrared (to 5400 nm) and to upgrade the visible as new line data or better 
spectra become available, I need at least \$300,000/year in grants
to the Smithsonian Astrophysical Observatory so that I can be rehired.
\vs\vs
\ni References

\ni Anderson, G. et al 1986. AFGL-TR-86-0110.

\ni Kurucz, R.L. 1992.  Model atmospheres for population synthesis.  Presented at 

IAU Symp. 149, Angra dos Reis, Brazil, August 5-9.  in Stellar Population 

of Galaxies, (eds. B. Barbuy and A. Renzini) Kluwer, Dordrecht, pp. 225-232.

\ni Kurucz, R.L., Furenlid, I., Brault, J., and  Testerman. L. 1984. {\it Solar Flux}

{\it Atlas from 296 to 1300nm}.  National Solar Observatory, Sunspot, New Mexico, 

240 pp.

\ni Thuillier, G., Floyd, L., Woods, T.N., Cebula, R., Hilsenrath, E., Herse, M., and

Labs, D. 2004.  Solar irradiance reference spectra. in {\it Solar Variability and its} 

{\it Effect on the Earth's Atmosphere and Climate System}, eds. J.M. Pap et al 

AGU, Washington, DC, pp. 171-194.

\vfill
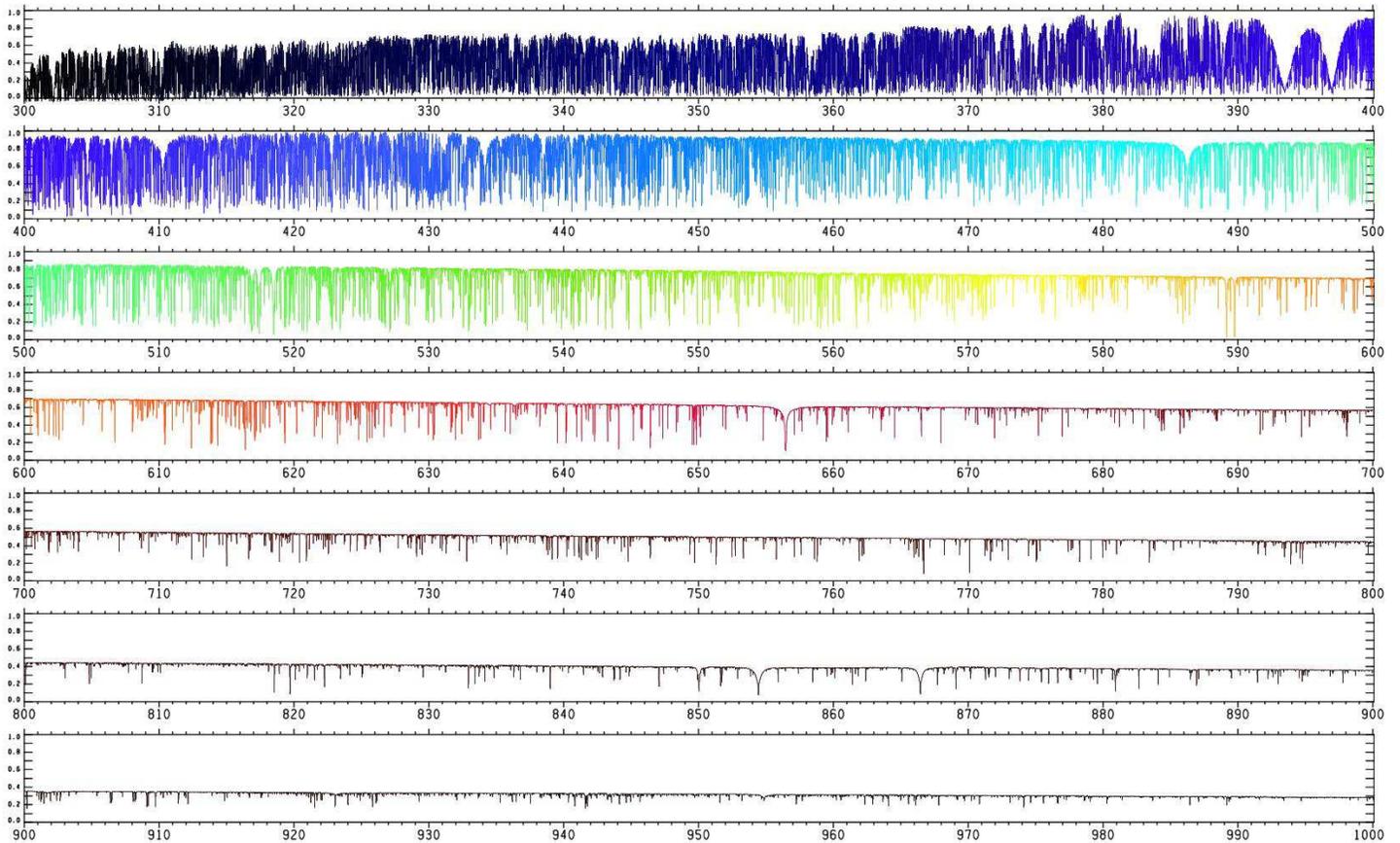
\end